\documentstyle[aps,prl,twocolumn]{revtex}
\input{epsf}

\begin{document}
\twocolumn[\hsize\textwidth\columnwidth\hsize\csname@twocolumnfalse\endcsname

\title{Hidden Chaos}
\author{Mikhail M.Sushchik and Nikolai F.Rulkov}
\address{
Institute for Nonlinear Science, University of California,
San Diego, La Jolla, CA 92093-0402\\
}

\maketitle

\begin{abstract} When a medium composed of microscopic elements is subjected to
a high intensity field, the individual behaviors of microscopic elements can
become chaotic.  In such cases it is important to consider the effects of this
irregularity at microscopical level onto the macroscopic behavior of the
medium.  We show that the macroscopic field produced by a large group of
chaotic scatterers can remain regular, due to the partial or
complete phase coherence of the scattering elements and the incoherence of the
chaotic components of their responses. Thus when
only macroscopic fields are observed,  one may be unaware of chaotic
microscopical motion, as it appears to be hidden from the  observer. The
coupling among the elements may lead to partial chaos synchronization, which
exposes the chaotic nature of the system making the oscillations of
macroscopic fields more irregular.
\end{abstract}
\narrowtext \vskip1pc]

The problem of wave scattering and dispersion is one of the basic problems in
many branches of physics. In most physical problems scattering and dispersion
are explained by the interaction of the wave field with particles that respond
linearly or weakly nonlinearly to the external forcing. However over the last
few years problems appeared where external fields are so powerful that they
drive the individual scatterers into the chaotic regime.  Examples of such
problems are the chaotic excitation of atoms and molecules by powerful
electro-magnetic  fields \cite{lasers}
and chaotic pulsations of cavitating bubbles in powerful ultrasonic fields
\cite{bubbles}.
In such cases the problem of scattering and dispersion becomes more
complicated. Even if the behavior of a microscopic element is chaotic, it is
not obvious what manifestation this would have at the macroscopic level. The
study of this issue is important from both theoretical and
experimental prospectives.

Scattering elements can typically be considered as passive damped oscillators.
When the amplitude of the scattered field is small compared to the external
field, the effect of the scattered field upon an individual element is
negligible and the whole system becomes an ensemble of uncoupled damped
oscillators driven by the same external force. Beyond some amplitude level the
nonlinearity of the scatterers becomes essential and can lead to chaotization
of their oscillations. In many cases chaotic oscillations occurring in a single
damped oscillator under the action of a periodic external force turn out to be
in some sense phase
locked to the driving force\cite{Rosenblum96}.
As the result, chaotic
oscillations in such systems contain components that are coherent with the
external driving.  If the macroscopic field is produced by a large group of
such microscopic elements, these components add coherently, while the chaotic
components of signals add non-coherently. The resulting macroscopic field
becomes nearly periodic.

When the coupling through the scattered field is taken
into account, an unusual situation occurs. Sufficiently strong coupling can lead
to partial or complete synchronization of chaotic oscillations in individual
elements
\cite{synchron}.
In such case chaotic components of these oscillations begin to add coherently
and the macroscopic field becomes more chaotic. Thus the synchronization
phenomenon, which is normally associated with the onset of a more regular
behavior in the system,  has the opposite effect on the observed macroscopic
quantities,  making their oscillations less regular.

To better understand this phenomenon, let us consider as an example the
following system:
\begin{equation}
\label{oscillator}
\frac{d^{2}x_j}{dt^{2}}+
\nu \frac{dx_j}{dt}-\frac{\alpha }{x_j^{3}}
+\frac{1}{x_j^{2}}=A\sin (\Omega t)+ \sum_{i=1}^N \frac{\kappa_{i,j}}{N}\frac{dx_i}{dt}.\nonumber
\end{equation}

These equations describe damped motion of $N$ particles in the potential \(
U(x)=\alpha /(2x^{2})-1/x \) and subject to the periodic external force. The
last term is to account for possible synchronization effects due to the mutual
coupling through the macroscopic field. Here we assume for simplicity
that the scattering region is much smaller than the  wavelength of the external
field and neglect the delays in the coupling term. The values
of the coupling coefficients depend on the geometry of the problem. We consider
the case where the scattered field is proportional to the derivative
of the state of a scattering element, $y_j(t)=dx_j(t)/dt$.  In the following discussion we shall
keep \( \nu =0.4 \), \( \alpha =0.75 \), \( A=0.45 \), and \( \Omega =1.3 \).

\begin{figure}[h]
\centerline{\epsfxsize=0.65\hsize \epsfbox{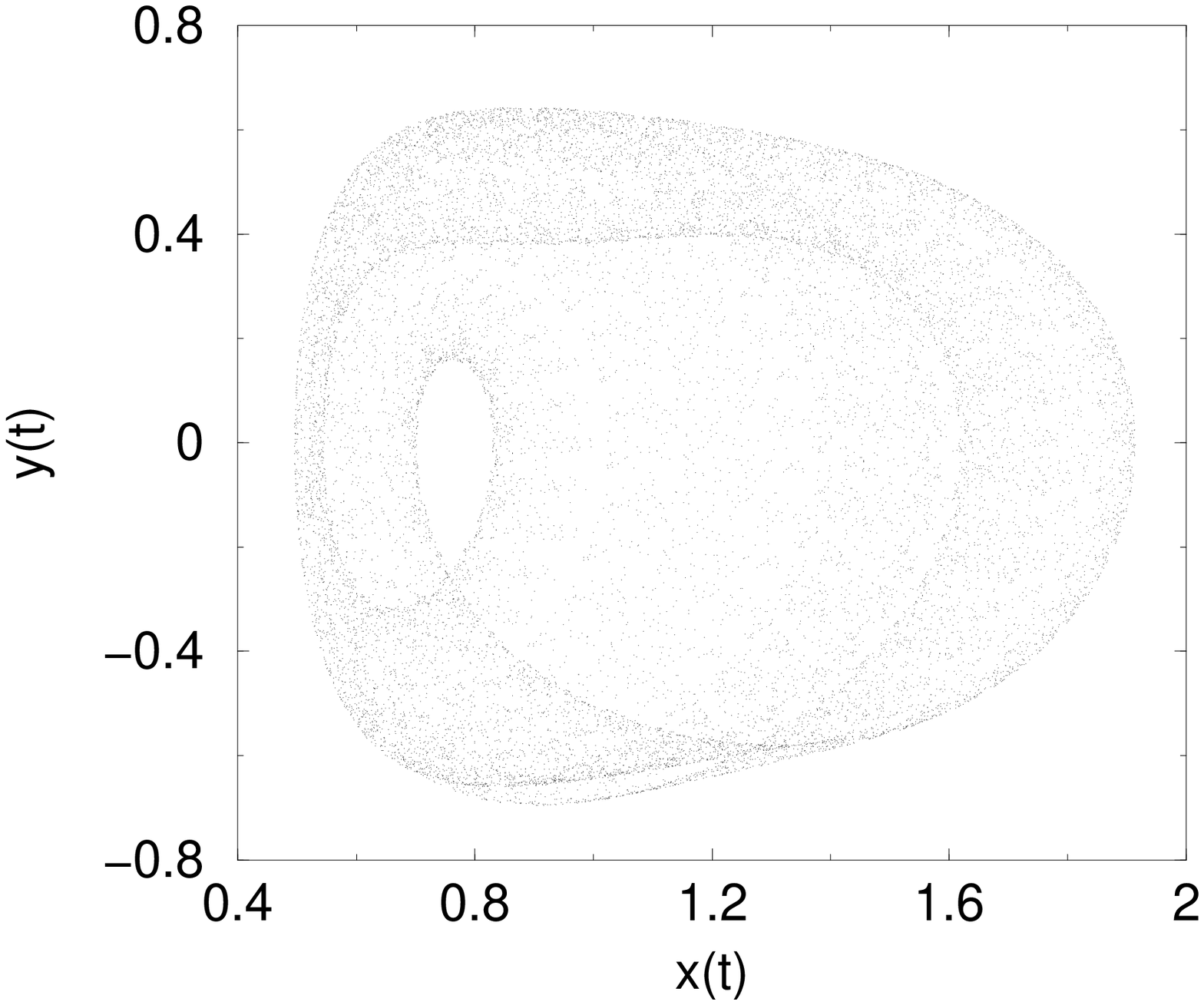}}
\caption{The chaotic attractor in one oscillator (\ref{oscillator}) without
coupling.\label{chaos1}}
\end{figure}

With these values of parameters and with $\kappa_{i,j}=0$ for all $i$ and $j$,
system (\ref{oscillator}) oscillates in a chaotic
regime, as illustrated in Fig.\ref{chaos1}.

The response of the ensemble, \( Y(t) \), observed
far from the scattering region is proportional to the sum of fields from
individual scatterers:

\[
Y(t)\sim \sum _{j=1}^{N}y_{j}(t).\]

Assuming stationarity of \( y_j(t) \), the mean temporal power spectral density
\( P(\omega ) \) of the scattered field can be calculated using \cite{Rytov87inls}

\[
P(\omega )=\frac{2}{\pi }\int _{0}^{\infty }C(\tau )\: cos\omega \tau \: d\tau ,\]
 where \( C(\tau ) \) is the auto-correlation function for \( Y(t) \):

\[
C(\tau )=<Y(t)Y(t+\tau )>
=\sum ^{N}_{j,k=1}<y_{j}(t)y_{k}(t+\tau )>\]

The last sum contains auto-correlation terms (\( j=k \)) and cross-correlation
terms (\( j\neq k \)).

Let us first consider the case when $\kappa_{i,j}=0$ for all $i$ and $j$. Then
all oscillators (\ref{oscillator}) evolve on the same attractor so that  \(
<y_{j}(t)y_{j}(t+\tau )>=C_{0}(\tau ) \), and \( <y_{j}(t)y_{k}(t+\tau
)>=C_{X}(\tau ) \) for $i\neq j$, independent of \( j \) and \( k \). Thus the
autocorrelation of the macroscopic field can be written as

\[
C(\tau )=N\left(C_{0}(\tau )-C_X(\tau)\right)+N^2C_{X}(\tau ),\]
and the power spectral density can be written as

\begin{equation}
\label{spectrum}
P(\omega )=N\left(P_{1}(\omega )-P_X(\omega)\right)+N^2P_{X}(\omega ),
\end{equation}
where \( P_{1}(\omega ) \) is the power spectral density of a single response,
and \( P_{X}(\omega ) \) is the cross spectral density.

\begin{figure}
\centerline{\epsfxsize=0.6\hsize \epsfbox{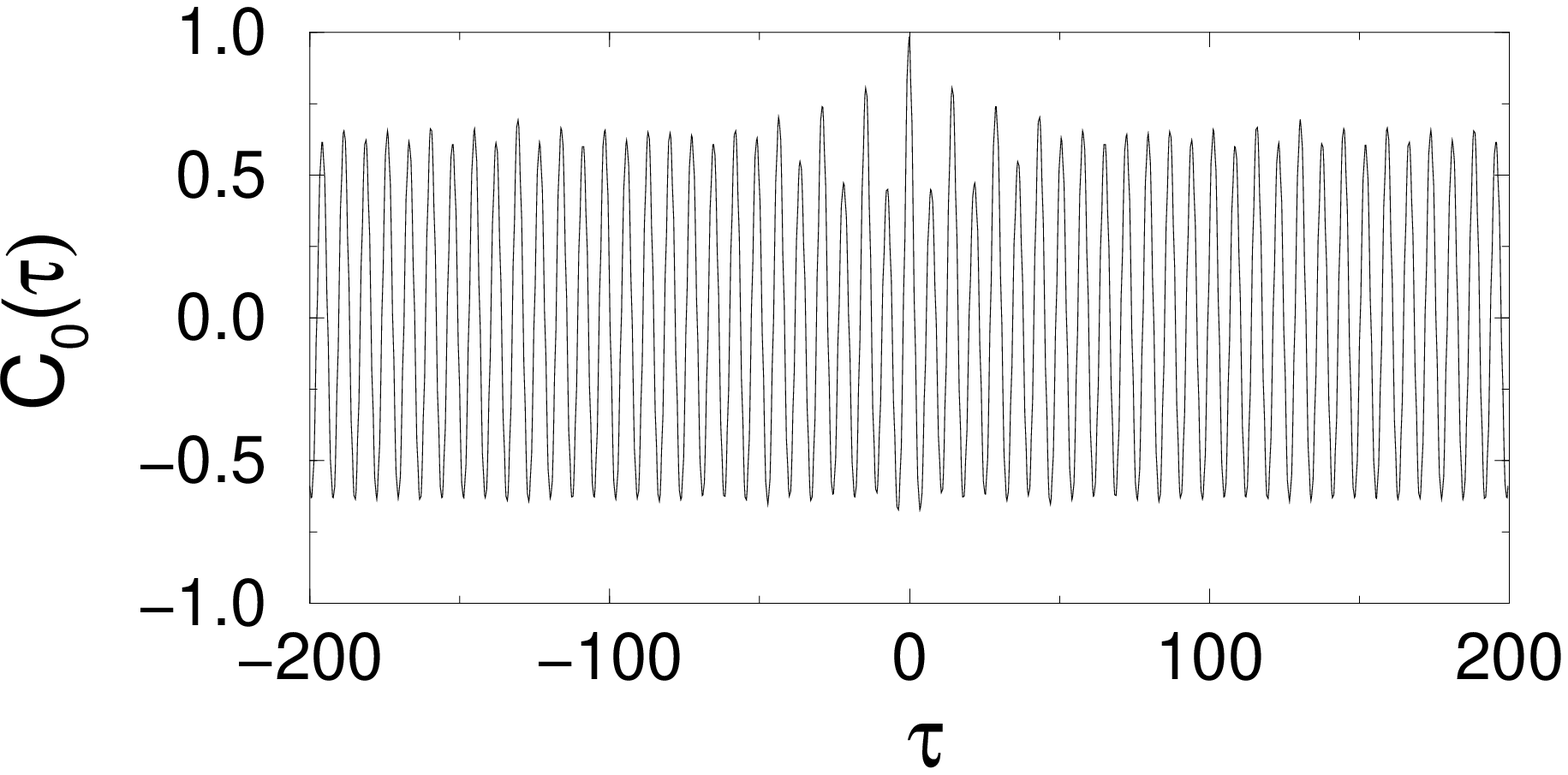}}
\centerline{\epsfxsize=0.6\hsize \epsfbox{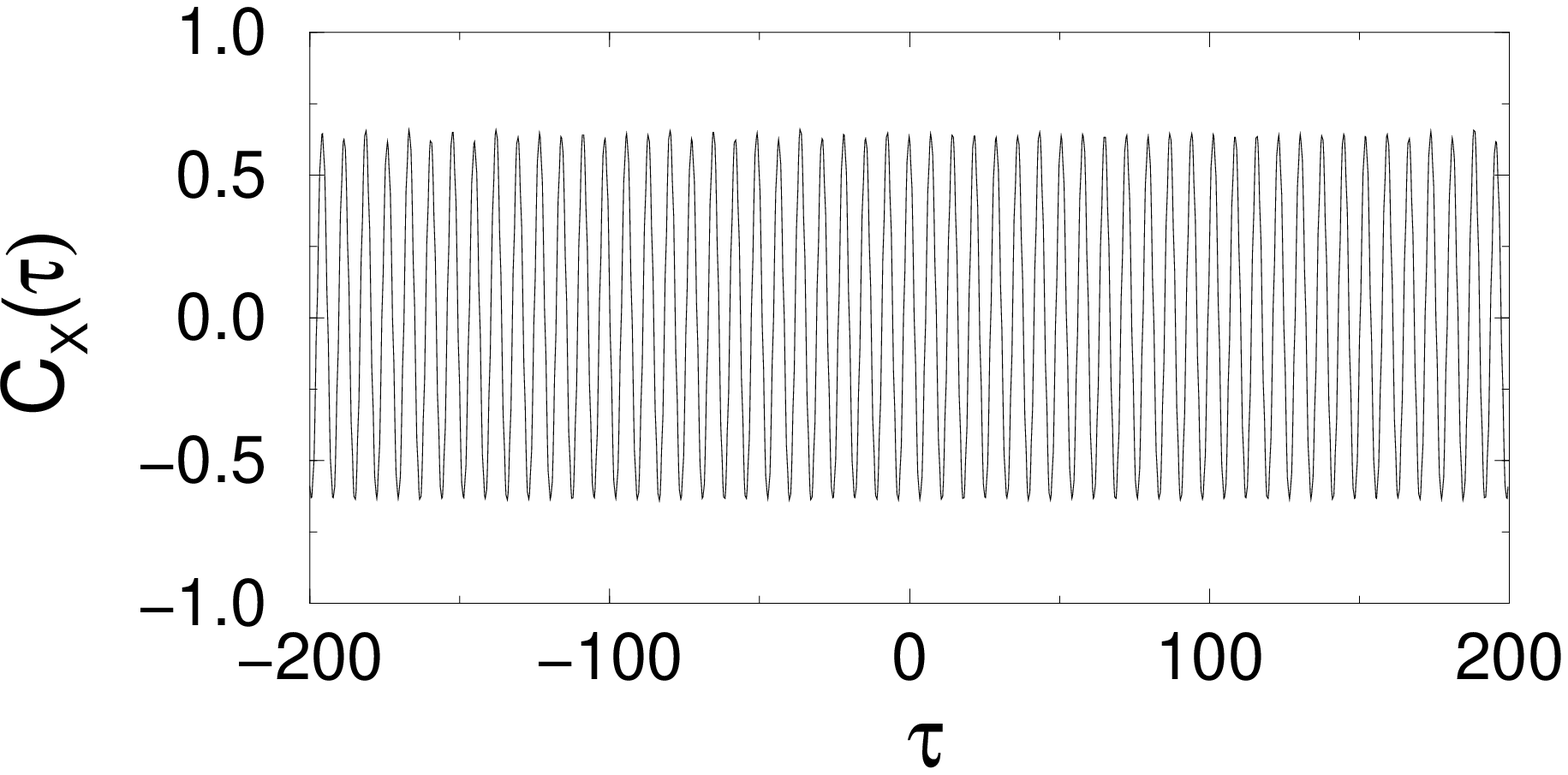}}
\centerline{\epsfxsize=0.6\hsize \epsfbox{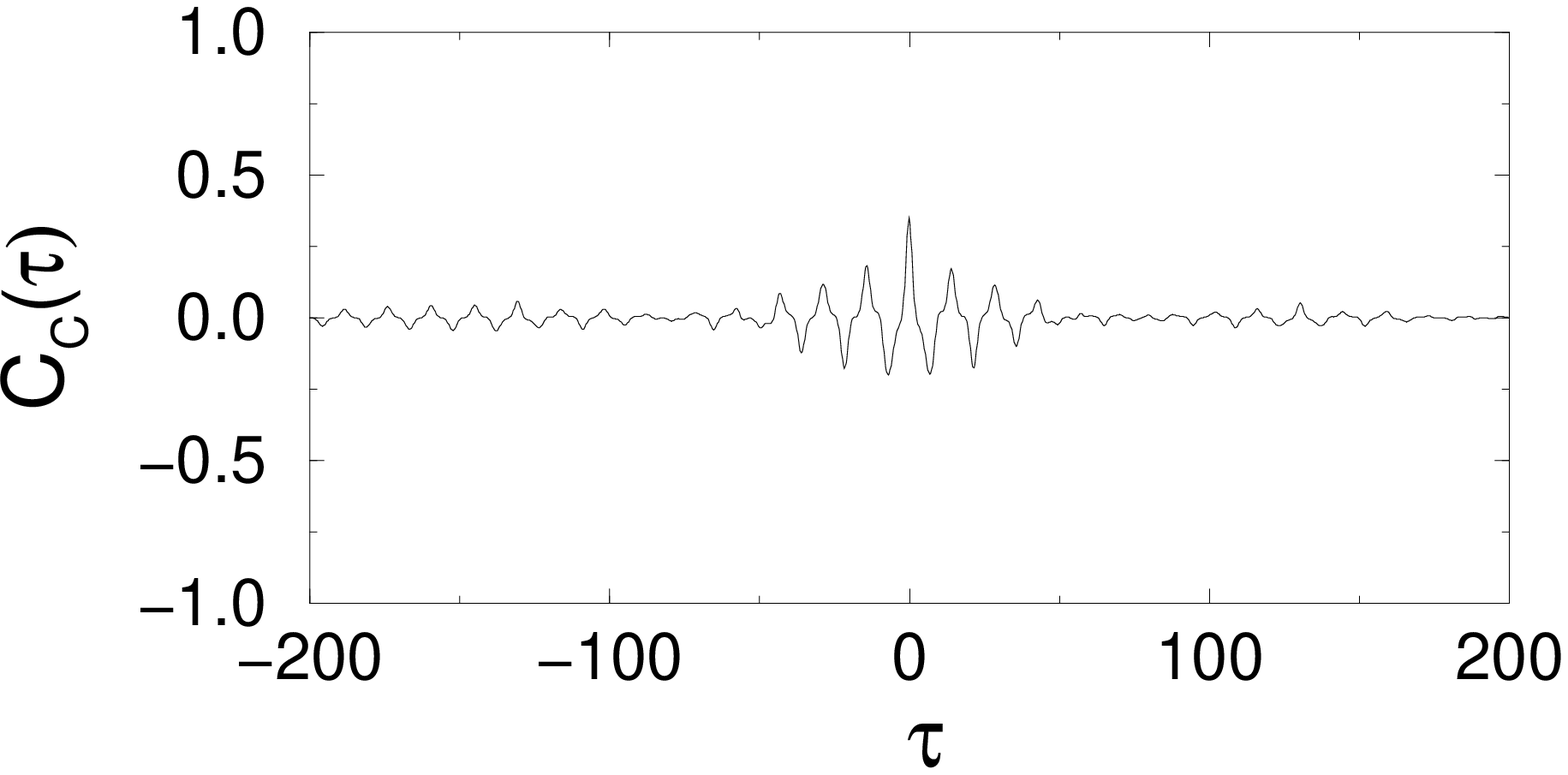}}
\caption{Autocorrelation of the chaotic response of oscillator
(\ref{oscillator}) without coupling (top); cross-correlation of responses of
two oscillators (middle); and the difference between these correlations
(bottom).\label{correl}}
\end{figure}

The autocorrelation, $C_0(\tau)$, the cross-correlation $C_X(\tau)$ and the
difference between the two, $C_C(\tau)=C_0(\tau)-C_X(\tau)$ are shown in
Fig.\ref{correl}. We see that $C_C(\tau)$ decays at large $|\tau|$, which means
that $P_{1}(\omega )-P_X(\omega)$ is the continuous component of the power
spectrum. More interestingly, we observe that $C_X(\tau)$ is a purely periodic
function, meaning $P_X(\omega)$ is the discrete component of the spectrum.

\begin{figure}[htb]
\centerline{\epsfxsize=\hsize  \epsfbox{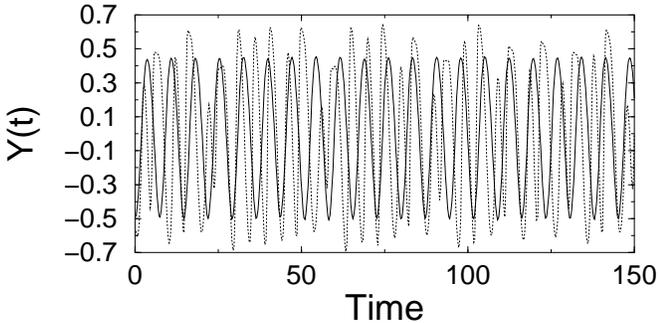}}
\caption{The mean field response of 4096 oscillators. The response of a single
element is shown with a dotted line for comparison.\label{macrofield}}
\end{figure}

\begin{figure}[htb]
\centerline{\epsfxsize=\hsize \epsfbox{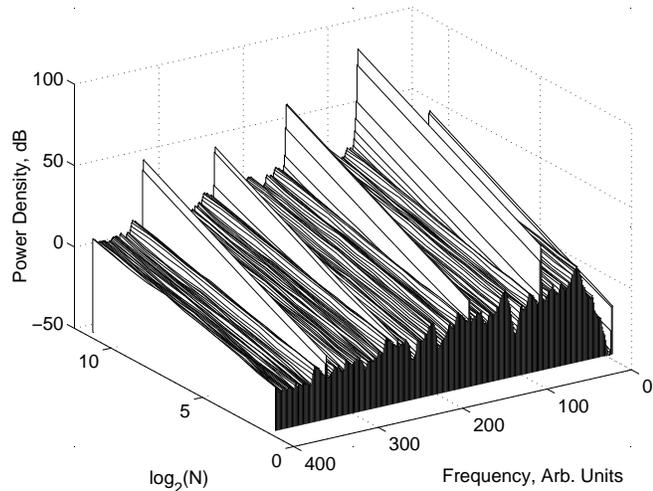}}
\caption{The mean power spectral density of the macroscopic response as a function of
the number of oscillators (in logarithmic scale).\label{power}}
\end{figure}

Equation \ref{spectrum}
indicates that as the number of scattering elements increases, the coherent
component of the spectrum increases quadratically with the number of elements,
while the continuous spectrum component increases linearly, as it typically
happens with non-coherent signals. Therefore, with a large number of chaotic
microscopic elements, the macroscopic response of the system is highly regular,
nearly periodic. Thus the chaotic nature of the system is hidden from the
macroscopic observer. This is illustrated in Fig.\ref{macrofield} and
Fig.\ref{power}. The faster growth of the discrete spectral component, compared
to the continuous chaotic component, is evident in Fig.\ref{power}.  A closer
look at a proper projection of this figure reveals good agreement with the
prediction of (\ref{spectrum}): the slope for the discrete part of the
spectrum is twice larger than for the continuous part.

In order to explain the mentioned features of the correlation functions,
let us study the phase properties of individual responses.
System (\ref{oscillator}) with zero coupling belongs to the class of
two-dimensional systems with periodic forcing, ${d\bf x}/dt={\bf F}({\bf
x},t)$, where ${\bf x} \in \Re^2$ and ${\bf F}({\bf x},t)={\bf F}({\bf
x},t+T)$, with strictly negative divergence of the vector field $\bf x$:
$\mbox{div}{\bf F}({\bf x},t)<0$ for any $\bf x$. By Liouville's formula it can
be shown that for such systems the Poincare map ${\bf
x}(t) \rightarrow {\bf x}(t+T)$ is area-contracting and therefore cannot
contain invariant circles\cite{math}. For the non-autonomous system ${\bf x}(t)$ this
means that it cannot posses quasiperiodic motions corresponding to trajectories
on invariant tori.  All motions in this system are either periodic motions
phase locked with the external forcing or chaotic motions.   If present,
chaotic trajectories will wonder in the vicinity of the skeleton composed of
periodic orbits, which are phase locked to the driving
force\cite{perorb}. The effect of this phase locking can be seen if we introduce
the phase of chaotic oscillations as the angle variable in the
angle-amplitude representation of the Hilbert transform, $h_j(t)$ \cite{Born64inls}:
$$
h_j(t)=y_j(t)+\frac{i}{\pi}\mbox{P.V.} \int_{-\infty}^\infty  \frac{y_j(\tau)}{\tau-t}d\tau,
$$

\begin{figure}
\centerline{\epsfxsize=0.7\hsize \epsfbox{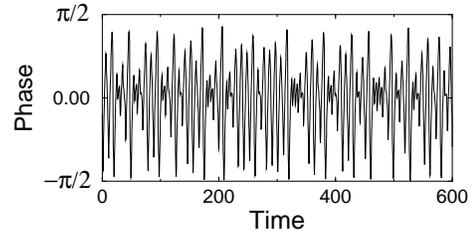}}
\caption{The detrended phase of the response of an individual oscillator.
\label{hilbert}}
\end{figure}

The phase mismatch between the response and the driving force,
$\Phi_j(t)=\phi_j(t)-\Omega t$, is shown in Fig.\ref{hilbert}. We observe that
\( \max (\Phi_j (t))-\min_j (\Phi (t))<\pi\). Thus the chaotic response oscillations
are loosely phase locked with the periodic driving and the fluctuations of the
relative phase are less than half-a-period. Due to this, the cross-correlation
of two oscillators driven by the same external field is purely periodic because
the effect of chaos averages to zero, while the
component coherent with the periodic driving does not. When $\tau$ is large,
due to the divergence of trajectories in the phase space of chaotic systems
and the loss of information about the initial condition,
the difference between computing autocorrelation for a single oscillator and
computing cross-correlation for signals from two different oscillators disappears.
This explains why $C_0(\tau)$ becomes purely periodic, and $C_C(\tau)$ decays to
zero, see Fig.\ref{correl}.

Let us now consider what happens when the coupling among the elements is taken
into account. For simplicity we assume that $\kappa_{i,j}=K$. Then the system (\ref{oscillator}) has the solution where $x_i(t)=X(t)$
for all $i$, which corresponds to identically synchronized chaotic oscillations
in the ensemble. If the coupling among the elements is sufficiently strong to
stabilize this solution, clearly, the mean field response is proportional
to that of a single element and is chaotic. Thus, although as synchronization
sets in the dynamics of the system becomes less chaotic, with a lower dimension
of the attractor, the mean field response becomes more irregular.

\begin{figure}
\centerline{\epsfxsize=0.7\hsize \epsfbox{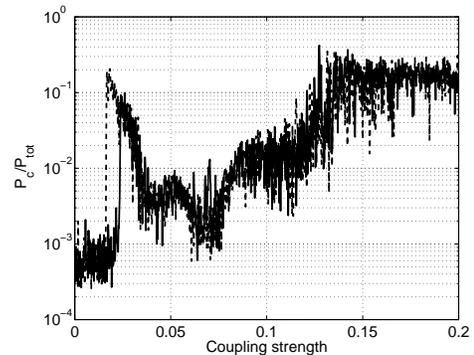}}
\caption{The ratio of the power in the continuous component of the spectrum to
the total power. The solid line corresponds to increasing the
coupling coefficient from zero, the dashed line, to decreasing
it.\label{hist}}
\end{figure}

Not quite so obvious is the effect of weak synchronizing coupling. When the
coupling is weak, the state of complete synchronization may not be achieved. To
make the matter more complicated, the variations of coupling lead to
changing dynamics of the entire systems. In addition to chaotic attractors
for some values of coupling periodic or quasiperiodic stated can become stable.
Nevertheless, the de-regularizing effects can be seen even for very small values
of the coupling. This is illustrated in Fig.\ref{hist} which shows the ratio of the power in the continuous
component of the spectrum to the total power as a function of coupling.
We see that for weak coupling we observe a sharp transition from a very regular
state where most of the power is in the periodic components to a less regular
state with a significant power in the continuous part of the spectrum. A
hysteresis is observed near the transition point. As the coupling increases
further, the dynamics of the system changes and the mean field response becomes
more regular again. At larger values of coupling, $\sim 0.2$, the system begins
to approach completely synchronized state. At $K=0.2$ the cross-correlation of
of responses of two oscillators looks almost identical to the autocorrelation of
a single oscillator, and the oscillations become quite irregular.

\begin{figure}[htb]
\centerline{\epsfxsize=\hsize  \epsfbox{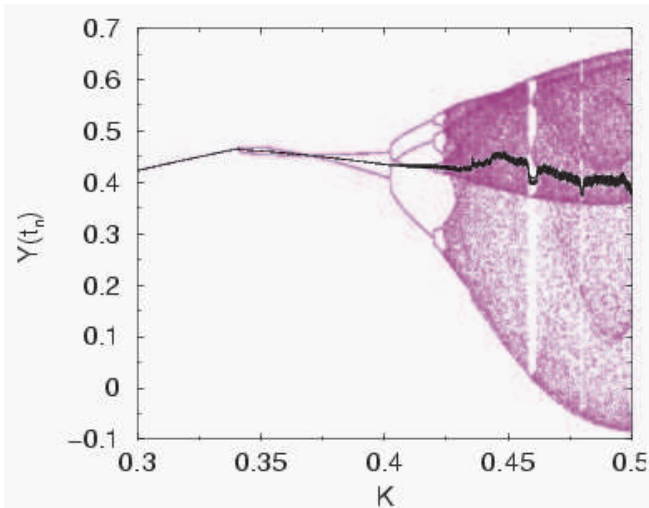}}
\caption{Bifurcation diagrams for a single oscillator and for the mean field
of 4086 elements (bold line). The diagram was created using the
Poincare map with the period of the driving.\label{bifurc}}
\end{figure}

In conclusion, we showed that when the only observed quantities are obtained by
averaging the responses from many chaotic oscillators to external fields,
these quantities can remain periodic. This presents a difficulty in observing
chaos in such systems. Fig.\ref{bifurc} illustrates the effect of averaging
onto the bifurcation diagram of our example system. We see that the period
doubling cascade and the chaotic regime, evident in the  bifurcation diagram of
a single element, are hardly visible in the bifurcation diagram of the mean
field. Under experimental conditions the bifurcation sequence and the
transition to chaos can easily be obscured by measurement noise\cite{newbub}.  The
macroscopic chaotic oscillations may arise in such systems as a result of
mutual coupling  between the elements. When this coupling is strong chaotic
oscillations in individual elements may become synchronized, which causes
non-periodic components of individual responses add coherently. As the
coupling strength varied, the transition
between regimes characterized by different degrees of regularity of the
mean field response can occur in a non-trivial way, for example, exhibit
hysteresis.

Partial coherence and effects similar to those discussed in this
communication can also occur in ensembles of mean field coupled
chaotic generators. In such systems  the external field may not be
necessary  to achieve phase coherence, which can arise
spontaneously due to a global mean field coupling in the
system\cite{Pikovsky96}.

We thank Lev Tsimring and Ulrich Parlitz for fruitful discussions. This research was supported in part by ARO, grant No.DAAG55-98-1-0269 and in part by the US
DOE, grant No. DE-FG03-95ER14516.

\bibliographystyle{prsty}

\end{document}